\documentclass[english,doublecol]{epl2}
\usepackage[T1]{fontenc}
\usepackage[latin1]{inputenc}
\usepackage{amsmath}
\usepackage{graphicx}

\makeatletter

\author{S. Mechkov\inst{1,2}, G. Oshanin\inst{1}, M. Rauscher\inst{3,4}\thanks{\email{rauscher@mf.mpg.de}}, M. Brinkmann\inst{5}, A. M. Cazabat\inst{2} and S. Dietrich\inst{3,4}}
\shortauthor{S. Mechkov \etal}
\institute{
\inst{1} Laboratoire de Physique Théorique de la Matière Condensée, Université Pierre et Marie Curie, Tour 24, Boîte 121, 4 place Jussieu, F-75252 Paris 05, France\\
\inst{2} Laboratoire de Physique Statistique de l'Ecole Normale
Supérieure, 24 rue Lhomond, F-75231 Paris 05, France\\
\inst{3} Max-Planck-Institut für Metallforschung, Heisenbergstr. 3, D-70569 Stuttgart, Germany\\
\inst{4} Institut für Theoretische und Angewandte Physik, Universität Stuttgart, Pfaffenwaldring 57, D-70569 Stuttgart, Germany\\
\inst{5} Max-Planck-Institut für Dynamik und Selbstorganisation, Bunsenstr. 10, D-37073 Göttingen, Germany
}
\abstract{Within the framework of a semi-microscopic interface displacement model we analyze the linear stability of sessile ridges and drops of a non-volatile liquid on a homogeneous, partially wet substrate, for both signs and arbitrary amplitudes of the three-phase contact line tension. Focusing on perturbations which correspond to deformations of the three-phase contact line, we find that drops are generally stable while ridges are subject only to the long-wavelength Rayleigh-Plateau instability leading to a breakup into droplets, in contrast to the predictions of capillary models which take line tension into account. We argue that the short-wavelength instabilities predicted within the framework of the latter macroscopic capillary theory occur outside its range of validity and thus are spurious.}


\makeatother

\usepackage{babel}
\makeatother
\begin{document}

\title{Contact line stability of ridges and drops}

\maketitle

\section{Introduction}

Many practically relevant processes involve liquids covering solid
substrates. The liquid may be a paint, a lubricant, an ink, or a dye.
The solid may be hard as a metal or soft as human skin or gel; it
may also show a homogeneous surface or be finely divided, as it is
the case for suspensions, porous media, or fibers. Due to the complexity
of these systems even extensive research in the past has still left
many open questions concerning in particular the intricate phenomena
taking place in the proximity of contact lines formed at three-phase
boundaries \cite{degennes85,degennes92,leger92,gbq}.

Considering a small liquid droplet deposited on a flat solid surface,
two distinct types of behavior can be observed depending on the sign
of the so-called spreading coefficient $S=\sigma_{SG}-\sigma_{SL}-\sigma$,
where $\sigma_{SG}$, $\sigma_{SL}$, and $\sigma$ are the tensions
of the three interphases which meet at the contact line: solid-gas,
solid-liquid, and liquid-gas, respectively. If, for a nonvolatile
liquid, $S\geq0$, the liquid tends to shield the solid from the vapor
phase which corresponds to the \textit{complete} wetting situation;
the drop will thus spread, tending to cover as much of the solid surface
as possible. Conversely, $S<0$ corresponds to the \emph{partial}
wetting regime; the drop will then evolve at some rate (see, \emph{e.g.},
ref.~\cite{decon} and refs.~therein) towards an equilibrium shape
characterized by a local thickness profile $\bar{h}(\mathbf{r})$
as function of the lateral coordinates $\mathbf{r}$ and an equilibrium
contact angle $\theta_{0}$ at which the liquid-gas interface meets
the solid surface. If the drop is small enough for gravity effects
to be negligible, it assumes the shape of an ideal spherical cap.
Typical examples of partial wetting situations are droplets of dew
on leaves or water drops on greasy glass.

Two centuries ago Young \cite{young} considered the balance of local
surface forces at the edge of such a drop and deduced that the equilibrium
contact angle $\theta_{Y}$ must satisfy \begin{equation}
\cos\theta_{Y}=\frac{\sigma_{SG}-\sigma_{SL}}{\sigma}\,.\label{eq:Young}\end{equation}
 For a long time, eq.~(\ref{eq:Young}) served as a tool for inferring
surface tensions from measurements of contact angles $\theta_{Y}$.
On the other hand, it was also understood long ago that Young's equation
is strictly valid only for the somewhat idealized geometry of a liquid
{}``wedge'' - a situation in which bulk phases are of macroscopic
extent, the interfaces are flat, and the contact line is straight.
Thus, for axisymmetric drops of base radius $R$, a modified Young's
equation holds, which takes into account, in leading order, the finite
drop size: \begin{equation}
\cos\theta_{0}=\cos\theta_{Y}-\frac{\tau}{\sigma R}\,,\label{eq:LineTensionForDrops}\end{equation}
 where $\tau$ is the excess free energy of the solid-liquid-vapor
system per unit length of the contact line \cite{marmur83,schimmele07}.
As the one-dimensional analogue of the surface tension, $\tau$ is
usually referred to as the contact line (CL) tension.

Formally, $\theta_{0}$ in eq.~(\ref{eq:LineTensionForDrops}) reduces
to $\theta_{Y}$ given by eq.~(\ref{eq:Young}) only in the limit
$R=\infty$ (straight CL) or if $\tau\equiv0$. Typical values of
$\tau$ are of the order of $10^{-10}$ to $10^{-11}$ J/m (see, \emph{e.g.},
refs.~\cite{marmur97,getta98,pompe00,bauer99,mora03,feijter72} and
refs.~therein) and may entail only small deviations of $\theta_{0}$
from $\theta_{Y}$ for generic values of $R$. However, there might
be \emph{other} interesting phenomena caused by the very fact that
$\tau$ is nonzero. We note that whereas surface tensions are necessarily
positive, $\tau$ is not constrained to be positive and may adopt
negative values as well \cite{gibbs61,bauer99}.

The most obvious consequence of a negative CL tension is that sufficiently
small drops no longer tend to coalesce, as soon as the negative CL
tension overcompensates the ensuing gain of surface free energy. This
critical drop size is of the order of $\tau/\sigma$, which is typically
in the nanometric range. Over the past few years, this nanoscale has
become directly accessible to experimental observation, which has
renewed the interest in negative CL tensions and their implications
\cite{brinkmann04,brinkmann05,dobbs99,marmur83,marmur97,rosso04,rosso06,indekeu92,indekeu94}.

A central issue of these studies is that of \emph{CL stability}. An
early analysis has been performed by Dobbs \cite{dobbs99} within
the framework of the interface displacement model (see, \emph{e.g.},
ref.~\cite{indekeu94}). Considering the ideal wedge geometry, he
has shown that a negative CL tension does not entail a CL instability
with respect to small amplitude deformations of any wavelength.

The question of the CL stability has been further addressed for other,
experimentally more relevant geometries such as axisymmetric drops
\cite{rosso06} or long liquid filaments (ridges) \cite{rosso04,brinkmann05},
within the framework of a macroscopic approach based on the second
variation of the free energy composed of the interface and line contributions
\cite{brinkmann04}. The latter authors have obtained complete phase
diagrams as function of $\theta_{0}$, $\tau$, and the perturbation
wavelength $\lambda$, showing both stable and unstable regimes. As
opposed to the interface displacement model, the macroscopic approach
with negative CL tension yields a range of unstable short wavelength
modes, bound from above by the critical wavelength\begin{equation}
\lambda_{0}=\frac{\left|\tau\right|}{\sigma\sin^{2}\theta_{0}}\,.\label{eq:CriticalWavelength}\end{equation}
Remarkably in ref.~\cite{brinkmann05} a negative $\tau$ was found
to destabilize a liquid filament with respect to contact line deformations
of \emph{any} wavelength for a sufficiently small contact angle $\theta_{0}$.
As it stands this result queries the standard view that the spontaneous
break-up of a filament into a chain of droplets (known as \emph{varicose}
or \emph{Rayleigh-Plateau instability}) occurs only for wavelengths
large compared with the lateral size of the filament. Additionally,
in ref.~\cite{brinkmann05} a \emph{positive} CL tension was shown
to \emph{stabilize} the varicose in accordance with the observation
that the critical (\emph{i.e.}, marginally stable) wavelength increases
with $\tau$.

We note, however, that the analyses in refs.~\cite{rosso06,rosso04,brinkmann05}
treat $\sigma$, $\theta_{0}$, and $\tau$ as independent parameters,
whereas for an actual system they are linked on the microscopic level.
Consequently, realistic partial wetting situations correspond to a
certain subspace of the surprisingly rich phase diagrams obtained
by the authors, and one may expect a microscopic or semi-microscopic
analysis to clarify the picture provided by those macroscopic models.

This so far unnoticed contradictory state of the literature prompted
us to revisit the problem of CL stability for liquid ridges and drops
within the framework of the interface displacement model. Our aim
is to determine what kind of instabilities can be encountered in realistic
models and hence, which parts of the phase diagrams deduced in refs.~\cite{rosso06,rosso04,brinkmann05}
are physically plausible.

In the following we first introduce the interface displacement model,
together with general considerations concerning the stability analysis.
We then focus on specific geometries, \emph{i.e.}, ridges and drops,
analyzing both with respect to perturbations leading to a deformation
of the contact line. We summarize our results in a final section.

\section{Basic properties of the interface displacement model}

We consider a two-dimensional, homogeneous, and flat substrate completely
covered by a thin liquid film; $h=h(x,y)$ describes the local film
thickness (film interface displacement relative to the substrate)
at a point $(x,y)$. $\Phi(h)$ denotes the effective interface potential,
which incorporates the net effect of the substrate-liquid and liquid-liquid
interactions \cite{degennes85,dietrich91a}. We consider a general
form of such potentials which might exhibit several minima and maxima;
we only stipulate that $\Phi(h)$ vanishes for $h\to\infty$, is differentiable,
is bound from below, and has its deepest minimum at a microscopic
thickness $h=a\geq0$ so that $\Phi^{\prime}(a)=0$. A typical example
of such a potential is \begin{equation}
\Phi\left(h\right)=\frac{A}{12\pi h^{2}}\left[1-\frac{1}{4}\left(\frac{a}{h}\right)^{6}\right]\,,\label{eq:PhiLJ}\end{equation}
 which corresponds to Lennard-Jones pair interactions, disregarding
other correction terms to the leading term $\sim h^{-2}$ \cite{dietrich91a}.
$A<0$ implies partial wetting.

The free energy per surface area of a planar liquid film with prescribed
thickness $h$ is given by $\Omega(h)=\sigma_{SL}+\sigma+\Phi(h)$
so that $\sigma_{SG}=\underset{h}{\min}\,\Omega(h)=\sigma_{SL}+\sigma+\Phi(a)$,
\emph{i.e.}, $\Phi\left(a\right)=S$. Accordingly, $\Phi(a)<0$ corresponds
to partial wetting.

This description formally applies only to films of constant thickness.
For laterally varying thickness profiles $h\left(x,y\right)$ the
model relies on the assumption that the two-dimensional gradient $\nabla\equiv\hat{\mathbf{x}}\partial_{x}+\hat{\mathbf{y}}\partial_{y}$
(\emph{i.e.}, the slope) of the profile $h$ is small, \emph{i.e.},
$\left|\nabla h\right|\ll1$. In this case the flat film expression
of the interface potential $\Phi(h)$ is locally valid.

The effective interface Hamiltonian $\mathcal{E}$ of such a smoothly
varying thin film has the form \cite{indekeu94}

\begin{equation}
\mathcal{E}\left[h\right]=\underset{\mathcal{A}}{\iint}\mathrm{d}x\mathrm{d}y\left[\frac{1}{2}\sigma\left(\nabla h\right)^{2}+\Phi\left(h\right)\right]\,,\label{eq:Energy2D}\end{equation}
in which the surface tension $\sigma$ reduces to a scaling parameter
for $\Phi$. $\mathcal{A}$ is the surface area. An important assumption
is that the liquid reaches the whole substrate, so that the nonwet
surface is covered by a film with a microscopic but finite thickness
$a$.

The stationary profile $\bar{h}(x,y)$ minimizes the functional in
eq.~(\ref{eq:Energy2D}) under geometry-specific constraints. As
in any meso- or microscopic approach, there is actually a contact
line region within which the interface interpolates smoothly between
the characteristic features of $\bar{h}(x,y)$. In the case of the
liquid wedge geometry (invariant along the $y$ axis), the asymptotes
of $\bar{h}(x)$ intersect such that the contact angle $\theta_{0}$
($=\theta_{Y}$ in this case) is formed \cite{indekeu92}:\begin{subequations}\label{eq:ContactAngle}\begin{eqnarray}
\theta_{0} & \equiv & \arccos\left[1+\Phi\left(a\right)/\sigma\right]\\
\theta_{0} & \approx & \sqrt{-2\Phi\left(a\right)/\sigma}\,.\end{eqnarray}
\end{subequations}The expressions in eq.~(\ref{eq:ContactAngle}),
with the latter being valid for $\theta_{0}\ll1$, are consistent
with the general expression of Young's law. Furthermore, the CL tension
$\tau$ of a wedge is given by the de Feijter and Vrij formula, referring
to the excess tangential force acting on the contact line region \cite{feijter72}
or the excess energy localised in that region \cite{indekeu92}:\begin{multline}
\tau=\sigma\int\limits _{a}^{\infty}\left[\left.\frac{\mathrm{d}h}{\mathrm{d}x}\right|_{x=\bar{h}^{-1}(h)}-\theta_{0}\right]\mathrm{d}h\label{eq:ThinFilmLT}\\
=\sqrt{2\sigma}\int\limits _{a}^{\infty}\left[\sqrt{\Phi\left(h\right)-\Phi\left(a\right)}-\sqrt{-\Phi\left(a\right)}\right]\mathrm{d}h\,.\end{multline}
Here $\bar{h}(x)$ minimizes $\mathcal{E}$ {[}eq.~(\ref{eq:Energy2D})]
with the boundary conditions $\partial_{x}\bar{h}\left(x\rightarrow-\infty\right)=0$
and $\partial_{x}\bar{h}\left(x\rightarrow+\infty\right)=\theta_{0}$~.

As pointed out in refs.~\cite{indekeu92,feijter72}, the interface
displacement model allows for arbitrary values of the line tension
{[}eq.~(\ref{eq:ThinFilmLT})] by means of appropriately chosen $\Phi\left(h\right)$
(slow decay, large humps, \emph{etc}). However, in particular for
negative line tensions, there is a close relationship between the
contact angle $\theta_{0}$, the line tension $\tau$, and a characteristic
thickness $\ell_{0}\equiv\frac{1}{\theta_{0}}\int_{a}^{+\infty}\left[\theta_{0}-\left.\frac{\mathrm{d}h}{\mathrm{d}x}\right|_{x=\bar{h}^{-1}(h)}\right]\mathrm{d}h\,$,
related to the range over which the interface potential $\Phi$ influences
the shape of the contact line. As an example, for the effective interface
potential in eq.~(\ref{eq:PhiLJ}) $\tau$ is negative and $\ell_{0}\approx a$.

In terms of this effective range $\ell_{0}$ one has $\tau=-\sigma\theta_{0}\ell_{0}$
and the critical wavelength $\lambda_{0}$ of CL instabilities as
predicted by eq.~(\ref{eq:CriticalWavelength}) (see, \emph{e.g.},
ref.~\cite{brinkmann05}) is given by $\lambda_{0}=\ell_{0}/\theta_{0}$,
\emph{i.e.}, it is comparable with the lateral width of the contact
line region within which the shape of the liquid-vapor interface is
influenced by $\Phi$. Thus the capillary model eq.~(\ref{eq:CriticalWavelength})
is based on predicts an instability at a length scale for which its
use is rather questionable. This motivates and encourages us to investigate
the stability of ridges and droplets by using a more microscopic model
which is able to cover the range of wavelengths below $\ell_{0}/\theta_{0}$.

\section{Stability analysis within the interface displacement model}

For the effective Hamiltonian in eq.~(\ref{eq:Energy2D}) the energy
variation $\tilde{\mathcal{E}}\equiv\mathcal{E}\left[\bar{h}+\varepsilon\Psi\right]-\mathcal{E}\left[\bar{h}\right]$
due to a small perturbation $\varepsilon\Psi(x,y)$ around a stationary
profile $\bar{h}(x,y)$ is given, to second order in $\varepsilon\ll1$,
by the expression \begin{multline}
\tilde{\mathcal{E}}=\varepsilon\iint\mathrm{d}x\mathrm{d}y\left[\Phi^{\prime}\left(\bar{h}\right)\Psi+\sigma\left(\nabla\bar{h}\right)\cdot\left(\nabla\Psi\right)\right]\\
+\frac{1}{2}\varepsilon^{2}\iint\mathrm{d}x\mathrm{d}y\left[\Phi^{\prime\prime}\left(\bar{h}\right)\Psi^{2}+\sigma\left(\nabla\Psi\right)^{2}\right]\,,\label{eq:Variation2D}\end{multline}
with $\Phi^{\prime}(h)=\mathrm{d}\Phi/\mathrm{d}h$ and $\Phi^{\prime\prime}(h)=\mathrm{d}^{2}\Phi/\mathrm{d}h^{2}$.
The requirement that the first variation vanishes for any perturbation
$\Psi$ preserving the volume of the liquid, \emph{i.e.}, with the
constraint\begin{equation}
\iint\Psi\mathrm{d}x\mathrm{d}y=0\,,\label{eq:VolumeConservationConstraint}\end{equation}
yields the Euler-Lagrange equation

\begin{equation}
\Phi^{\prime}\left(\bar{h}\right)-\sigma\nabla^{2}\bar{h}=P\label{eq:Stationary2D}\end{equation}
for the equilibrium profile $\bar{h}$. The Lagrange parameter $P$
fixes the volume and plays the role of the Laplace pressure so that
for the homogeneous equilibrium film thickness $\bar{h}(\mathbf{r})=a$
one has $P=0$.

The stability condition is that the second variation

\begin{equation}
\tilde{\mathcal{E}}=\frac{1}{2}\varepsilon^{2}\iint\mathrm{d}x\mathrm{d}y\left[\Phi^{\prime\prime}\left(\bar{h}\right)\Psi^{2}+\sigma\left(\nabla\Psi\right)^{2}\right]\label{eq:2ndVariation2D}\end{equation}
 is non-negative for any perturbation $\Psi$ fulfilling eq.~(\ref{eq:VolumeConservationConstraint}).


For a suitably normalized $\Psi$, the integral on the right-hand
side of eq.~(\ref{eq:2ndVariation2D}) is the expectation value of
the Hermitian operator\begin{eqnarray}
\hat{P} & = & U\left(x,y\right)-\sigma\nabla^{2}\label{eq:PressureOperator}\end{eqnarray}
 where $U\left(x,\, y\right)\equiv\Phi^{\prime\prime}\left[\bar{h}(x,y)\right]$.
We call $\hat{P}$ the {}``pressure operator'' because $\varepsilon\hat{P}\Psi$
is the additional local pressure generated by the perturbation $\varepsilon\Psi$
(\emph{cf} eq.~(\ref{eq:Stationary2D}) for stationary contributions).

Analogous to quantum mechanics, $\Psi$ can be regarded as the {}``wave
function'' of a particle and $\hat{P}$ as its {}``Hamiltonian''
with potential energy $U(x,y)$ and kinetic energy $\left(-\sigma\nabla^{2}\right)$.
Thus the linear stability analysis of the stationary profile $\bar{h}$
reduces to an eigenvalue problem for $\hat{P}$: \begin{equation}
E\Psi=\hat{P}\Psi\,,\label{eq:Schroedinger2D}\end{equation}
 which has the structure of the time-independent Schrödinger equation.
As it will turn out, the elementary perturbations of interest automatically
preserve the volume of ridges and drops, and thus in every specific
case the constraint in eq.~(\ref{eq:VolumeConservationConstraint})
is satisfied.

The eigenfunctions with positive energy $E$ are called stable, while
the ones with $E<0$ contribute to the instability of the system.
Thus the stationary solution $\bar{h}$ is only stable if $\hat{P}$
has no eigenfunctions with $E<0$.

A basic feature of the Schrödinger equation is that its ground state
has no zeros inside its domain of definition (see §~20 in ref.~\cite{LL3}
and chap.~6 in ref.~\cite{CH1}). This holds not only for the 2D
Hamiltonian $\hat{P}$, but also for the effective 1D Hamiltonian(s)
it reduces to for geometries with translational or rotational symmetry.

Another basic feature, valid for any stationary film structure on
a homogeneous substrate, is the indifference with respect to lateral
shifts in any direction $\hat{\mathbf{u}}$ of the $\left(x,y\right)$
plane, following from the translational symmetry on a homogeneous
substrate ($\Phi$ does not depend on $x$ or $y$ explicitly). Specifically,
for any unit vector $\hat{\mathbf{u}}$, $\Psi=\hat{\mathbf{u}}\cdot\nabla\bar{h}$
is an eigenfunction of $\hat{P}$ with eigenvalue $0$:

\begin{eqnarray}
\hat{P}\,\left(\mathbf{u}\cdot\nabla\bar{h}\right) & = & \Phi^{\prime\prime}\left(\bar{h}\right)\left(\hat{\mathbf{u}}\cdot\nabla\bar{h}\right)-\sigma\nabla^{2}\left(\hat{\mathbf{u}}\cdot\nabla\bar{h}\right)\nonumber \\
 & = & \hat{\mathbf{u}}\cdot\nabla\underbrace{\left[\Phi^{\prime}\left(\bar{h}\right)-\sigma\nabla^{2}\bar{h}\right]}_{P=const}=0\,.\label{eq:LateralShiftStability}\end{eqnarray}
The knowledge of the explicit expression $\Psi=\hat{\mathbf{u}}\cdot\nabla\bar{h}$
for lateral shift modes with zero energy turns out to be very useful
whenever the stationary solutions exhibit specific geometrical properties
(\emph{e.g.}, ridges or drops). This forms the core of our following
analysis, together with exploiting the existence of a unique ground
state with no zeros.

\section{Ridge stability: preliminaries}

If the liquid structure is translationally invariant along the $y$
axis, the stationary solutions are of the form $\bar{h}(x)$ and eq.~(\ref{eq:Stationary2D})
reduces to\begin{equation}
\Phi^{\prime}\left(\bar{h}\right)-\sigma\partial_{x}^{2}\bar{h}=P\,.\label{eq:Stationary1D}\end{equation}
The following analysis is based on solutions $\bar{h}(x)$ of eq.~(\ref{eq:Stationary1D})
which resemble a liquid \emph{ridge} centered symmetrically around
$x=0$ with $\bar{h}\left(\left|x\right|\rightarrow\infty\right)=h_{0}$
and $\partial_{x}\bar{h}\left(\left|x\right|\rightarrow\infty\right)=0$
so that $P=\Phi^{\prime}(h_{0})$. Thus the first integral of eq.~(\ref{eq:Stationary1D})
is \begin{equation}
\frac{1}{2}\sigma\left(\partial_{x}\bar{h}\right)^{2}=\Phi\left(\bar{h}\right)-\Phi\left(h_{0}\right)-\left(\bar{h}-h_{0}\right)P\,.\label{eq:Stationary1Dintegrated}\end{equation}
A typical solution of eq.~(\ref{eq:Stationary1Dintegrated}) is shown
in fig.~\ref{fig:EnergyWell} (solid curve).

In view of studying eqs.~(\ref{eq:PressureOperator}) and (\ref{eq:Schroedinger2D})
in the present case for which $U(x,y)=U(x)$, we can restrict the
linear stability analysis to the Fourier components of the deformations:
\begin{equation}
\Psi(x,y)=\psi(x;k)\cos\left[k\, y+\alpha(k)\right],\, k\neq0\,,\label{eq:Mode1D}\end{equation}
 with arbitrary phase shifts $\alpha(k)$. For $k\neq0$, $\Psi$
automatically satisfies the volume conservation constraint in eq.~(\ref{eq:VolumeConservationConstraint}).
Thus for $k\neq0$ eq.~(\ref{eq:Schroedinger2D}) reduces to the
1D Schrödinger equation:\begin{eqnarray}
\hat{H}\psi\left(x\right)=E\psi(x) & , & \hat{H}=\bar{H}+\sigma k^{2}\,,\label{eq:Schroedinger1D}\\
\bar{H}=U\left(x\right)-\sigma\partial_{x}^{2} & , & U\left(x\right)=\Phi^{\prime\prime}\left[\bar{h}\left(x\right)\right]\,.\label{eq:PseudoHamiltonian0}\end{eqnarray}
If $\psi$ is an eigenfunction of $\bar{H}$ with eigenvalue $\bar{E}$,
it is also an eigenfunction of $\hat{H}$ with eigenvalue $E=\bar{E}+\sigma k^{2}$.
Due to $\sigma>0$ this relation allows us to infer the stability
from the limiting case $k\rightarrow0$, which is the least stable
mode. Thus the 2D mode $\Psi(x,y)=\psi(x)\cos\left(k\, y\right)$
is stable ($E>0$) for all $k$, unless $\bar{E}<0$, in which case
there is an instability for $\left|k\right|<\left(-\bar{E}/\sigma\right)^{1/2}$.

With that at this point the general results are:

\begin{itemize}
\item Within the framework of the interface displacement model a stationary
solution of eq.~(\ref{eq:Stationary1D}) can be unstable only with
respect to long-wavelength perturbations (such as Rayleigh-Plateau
instabilities).
\item The discussion of unstable modes can be reduced to that of eigenfunctions
$\psi$ of $\bar{H}$ with eigenvalues $\bar{E}<0$. 
\end{itemize}
This reasoning, which does not depend on the form of $\Phi(h)$, rules
out linear instabilities of ridges with respect to short-wavelength
deformations of the contact line.

The following section provides a more detailed study of the long-wavelength
instability of ridges, including the limiting case of a wedge, which
can be viewed as a ridge of macroscopic transverse size. Finally we
shall turn our attention to axisymmetric drops.

\section{Ridge stability: detailed analysis}

\begin{figure}
\includegraphics[width=1\columnwidth,keepaspectratio]{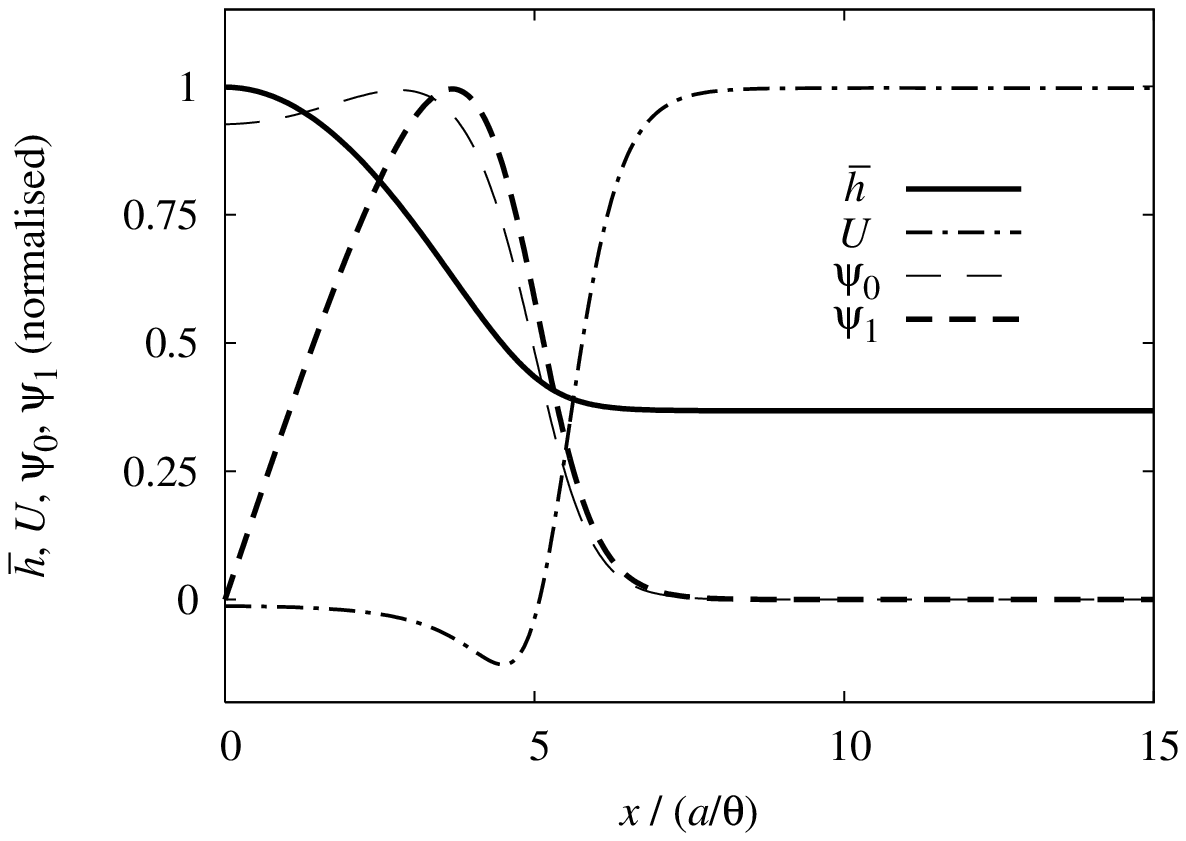}

\caption{\label{fig:EnergyWell}Stationary profile $\bar{h}\left(x\right)/\bar{h}\left(0\right)$
of a liquid ridge (with $\bar{h}\left(0\right)\simeq2.81\, a$ and
$h_{0}=1.04\, a$) and corresponding potential energy $U\left(x\right)=\Phi^{\prime\prime}\left[\bar{h}(x)\right]$
(eq.~(\ref{eq:PressureOperator})) of the 1D Hamiltonian, normalized
by $U(\infty)\simeq4.92\,\sigma\theta^{2}/a^{2}$. Also shown are
the two eigenstates of $\hat{P}$ of lowest energy, $\psi_{0}$ and
$\psi_{1}$, normalised by their maximal values. $\bar{h}$, $U$,
and $\psi_{0}$ are symmetric and $\psi_{1}$ is antisymmetric.}
\end{figure}

Figure \ref{fig:EnergyWell} shows the vertical cross-section along
$x$ of a ridge (solid curve) centered at $x=0$ and aligned with
the $y$ axis. The basic features are an apex at $x=0$ and a symmetric
decay on both sides towards $\bar{h}\left(-\infty\right)=\bar{h}\left(+\infty\right)=h_{0}$.
The thickness $h_{0}$ corresponds to a stable film, \emph{i.e.},
$\Phi^{\prime\prime}\left(h_{0}\right)>0$ so that $U\left(\pm\infty\right)>0$.
For eq.~(\ref{eq:PhiLJ}) this implies $h_{0}<\sqrt[6]{3}a$.

By analogy with quantum mechanics, an eigenstate $\psi$ of $\bar{H}$
with eigenvalue $\bar{E}<0$ is \emph{bound} by the potential $U(x)=\Phi^{\prime\prime}\left[\bar{h}\left(x\right)\right]$.
According to the so-called {}``oscillation theorem'' (see, \emph{e.g.},
§~21 in ref.~\cite{LL3}), the bound states are non-degenerate and
indexed by their number of nodes. Since the stationary solution $\bar{h}(x)$
and thus the potential energy $U(x)$ are symmetric with respect to
$x=0$, the eigenstates are either even (symmetric) or odd (antisymmetric).

As a special case of eq.~(\ref{eq:LateralShiftStability}), $\psi_{1}\left(x\right)\equiv\partial_{x}\bar{h}\left(x\right)$
is an eigenfunction of $\bar{H}$ with eigenvalue $\bar{E}_{1}=0$.
The associated 2D deformation [eq.~(\ref{eq:Mode1D})] corresponds,
for $k=0$, to a lateral shift of the ridge, and for $k\neq0$, to
finite-wavelength meandering. Since $\psi_{1}$ is an eigenmode of
$\hat{H}$ with eigenvalue $E_{1}=\sigma k^{2}$, a ridge is generally
stable with respect to meandering.

The eigenfunction $\psi_{1}$ has a single node at $x=0$, \emph{i.e.},
at the apex of the ridge. From the properties of bound states we infer
a unique ground state $\psi_{0}$ for $\bar{H}$ with an energy $\bar{E}_{0}<\bar{E}_{1}$.
The mode $\psi_{0}$ has no nodes: the associated 2D perturbation
corresponds, for $k\neq0$, to periodic bulging and shrinking of the
cross-section of the ridge, and is effectively unstable for $\left|k\right|<\left(-\bar{E}_{0}/\sigma\right)^{1/2}$
(varicose or Rayleigh-Plateau instability). Perturbations with wavelengths
shorter than $\bar{\lambda}_{0}=2\pi/\left(-\bar{E}_{0}/\sigma\right)^{1/2}$
are stable. As the only eigenmode of $\bar{H}$ with $\bar{E}<0$,
for any ridge $\psi_{0}$ is the only source of instability.%
Figure \ref{fig:EnergyWell} shows both $\psi_{1}$ and $\psi_{0}$
for a ridge, together with the stationary profile $\bar{h}$ and the
symmetric potential well $U\left(x\right)=\Phi^{\prime\prime}\left[\bar{h}\left(x\right)\right]$.
$\Phi$ is given by eq.~(\ref{eq:PhiLJ}), \emph{i.e.}, the line
tension is negative.

Unlike for $\psi_{1}$, there are no general analytic expressions
for $\psi_{0}$ and its energy $\bar{E}_{0}$. Reference~\cite{unpublished}
considers the limit of large ridge width $L_{x}=\frac{1}{\delta}4\, a/\theta_{0}$
and small internal pressure $P=\delta\,\Phi(a)/a$, where $\delta$
is a small dimensionless parameter. In that limit, it can be shown
\cite{explanations} that the marginally stable wavelength $\bar{\lambda}_{0}$
scales linearly with the ridge width $L_{x}$. The key result is that
the limiting ratio $\underset{L\rightarrow\infty}{\lim}\bar{\lambda}_{0}/L\simeq2.62$
is \emph{intrinsic}: it depends neither on details of $\Phi\left(h\right)$
nor on the value of $\tau$ as determined by eq.~(\ref{eq:ThinFilmLT}).
Interestingly, ref.~\cite{brinkmann05} also analyzes the limit $\theta_{0}\rightarrow0$
and reports $\bar{\lambda}_{0}/L\simeq2.62$ for the specific case
in which the CL tension, used as an input variable in that macroscopic
approach, is \emph{exactly zero}.

The linear scaling of the marginally stable wavelength with the ridge
width is consistent with the marginal stability of the \emph{wedge},
viewed as a ridge of macroscopic size with pressure $P=0$ and $h_{0}=a$
(see ref.~\cite{dobbs99} for a separate derivation of this stability
of the wedge considered on its own). The scaling also complies with
the intuitive picture of the Rayleigh-Plateau instability as the \emph{only}
mechanism through which a macroscopic ridge breaks up into a chain
of droplets, the size and spacing of the drops also being proportional
to the width of the ridge. In the following we shall demonstrate that
drops are not unstable with respect to deformations affecting the
circular shape of the CL.

\section{Stability of axially symmetric drops\label{sec:Drops}}

For the discussion of axisymmetric drops we introduce polar coordinates
$\left(r,\varphi\right)$. With a symmetric ansatz $\bar{h}(r)$ for
the stationary solution eq.~(\ref{eq:Stationary2D}) turns into\begin{equation}
\Phi^{\prime}\left(\bar{h}\right)-\sigma\frac{1}{r}\partial_{r}\left[r\,\partial_{r}\bar{h}(r)\right]=P\,.\label{eq:Stationary1DR}\end{equation}
Similarly to the $y$-invariant case, eq.~(\ref{eq:Stationary1DR})
admits solutions $\bar{h}(r)$ that have a single apex at $r=0$ (\emph{i.e.},
$\partial_{r}\bar{h}(r)=0$ for $r=0$) and tend to a thickness $h_{0}$
at $r=\infty$ with $\partial_{r}\bar{h}(r)<0$ for $r>0$. These
are the equilibrium droplet shapes on which we base our stability
analysis.

In view of the rotational symmetry of the basic solution of eq.~(\ref{eq:Stationary1DR})
we again focus on the Fourier components of the perturbations: \begin{equation}
\Psi(r,\varphi;m)=\psi(r;m)\cos\left[m\,\varphi+\alpha(m)\right]\,.\label{eq:Mode1DR}\end{equation}
 Equation (\ref{eq:Mode1DR}) is similar to eq.~(\ref{eq:Mode1D}),
with the important difference that the continuous wave vector $k$
is replaced by an integer $m$ (resembling an azimuthal quantum number).

We restrict our analysis to positive $m$, because the deformations
associated with $m=0$ do not affect the circular shape of the contact
line. For $m\geq1$, the volume conservation constraint in eq.~(\ref{eq:VolumeConservationConstraint})
is automatically satisfied, and the linear stability analysis reduces
to the following 1D eigenvalue problem for $\psi\left(r\right)$:\begin{eqnarray}
\hat{H}_{m}\psi(r)=E\psi(r) & , & \hat{H}_{m}=\hat{T}+\hat{U}_{m}\,,\label{eq:EigenvalueProblem1DR}\\
\hat{U}_{m}=U\left(r\right)+\sigma\frac{m^{2}}{r^{2}} & , & U\left(r\right)=\Phi^{\prime\prime}\left[\bar{h}\left(r\right)\right]\,,\\
\hat{T} & = & -\sigma\frac{1}{r}\partial_{r}\left(r\,\partial_{r}\right)\,.\end{eqnarray}

Instability is again indicated by the occurrence of negative eigenvalues
of $\hat{H}_{m}$. Before we address the overall stability of a stationary
profile $\bar{h}$, it is useful to make the following remarks about
$\hat{H}_{m}$:

\begin{enumerate}
\item With respect to the scalar product $\left\langle \psi\varphi\right\rangle =\int r\,\mathrm{d}r\,\psi^{*}\varphi$,
the operator $\hat{T}$ is Hermitian and positive definite.
\item For every value of $m$, $\hat{H}_{m}$ is a Hermitian 1D operator
consisting of a potential term $\hat{U}_{m}$ and a kinetic term $\hat{T}$.
The existence of a non-degenerate ground state with energy $E_{m}$
and without nodes (see, \emph{e.g.}, §~20 in ref.~\cite{LL3} and
chap.~6 in ref.~\cite{CH1}) holds for $\hat{H}_{m}$.
\item $E_{m}<E_{n}$ if $m<n$. Indeed, for any $\psi\left(r\right)$, one
has\begin{equation}
\left\langle \hat{H}_{n}\right\rangle _{\psi}=\left\langle \hat{H}_{m}\right\rangle _{\psi}+\sigma\left(n^{2}-m^{2}\right)\left\langle r^{-2}\right\rangle _{\psi}\,.\label{eq:rem3}\end{equation}
 Specifically, for $m<n$ and $\psi$ chosen to be the ground state
of $\hat{H}_{n}$, one has $E_{m}<\left\langle \hat{H}_{m}\right\rangle _{\psi}<\left\langle \hat{H}_{n}\right\rangle _{\psi}=E_{n}$.
\end{enumerate}
As in the case of the ridge, the drop is marginally stable with respect
to lateral shifts: \begin{eqnarray}
\psi_{1}\left(r\right)=\partial_{r}\bar{h}(r) & \Rightarrow & \hat{H}_{1}\psi_{1}=0\,.\end{eqnarray}
 The eigenfunction $\psi_{1}$ has no nodes other than the one at
$r=0$, and thus it is the ground state of $\hat{H}_{1}$; hence (see
remark 3) $E_{m}\geq0$ for all $m\geq1$. This allows us to affirm
that axially symmetric drops are not subject to instabilities which
are associated with corrugations of the contact line.

\section{Conclusion}

Within the framework of a semi-microscopic interface displacement
model we have analyzed the linear stability of stationary distributions
of a non-volatile liquid on a homogeneous, partially wet substrate.
We have examined macroscopically extended filaments (ridges) and axially
symmetric drops. The stability analysis has been focused on deformations
affecting the shape of the CL which without perturbation is straight
for ridges and circular for drops. Within our approach the CL tension
is borne out as a microscopic property associated with the intrinsic
morphology of the three-phase contact region, rather than being an
arbitrary macroscopic input parameter.

For each of the two aforementioned geometries, we have considered
the elementary perturbations which induce periodic deformations of
the CL. We have shown that these perturbations correspond to a positive
second variation of the free energy of the liquid structures, with
the sole exception of the expected Rayleigh-Plateau instability for
ridges. Notably, the critical wavelength of the Rayleigh-Plateau instability
is not affected by CL tension. Therefore the interface displacement
model leaves no room for short-wavelength CL instabilities. This stability
holds for generic liquid-substrate and liquid-liquid interactions
captured by the effective interface potential $\Phi$, which allows
for arbitrary values and either sign of the CL tension. This approach
also covers the limiting cases for which the CL tension is ill-defined
according to eq.~(\ref{eq:ThinFilmLT}): \emph{e.g.}, for a film
of nematic liquid crystals with antagonistic anchoring $\Phi$ includes
a long-ranged elastic term so that $\Phi(h)\sim1/h$ for $h\rightarrow\infty$
(see, \emph{e.g.}, ref.~\cite{valignat96}).

We conclude that the short wavelength instability found in ref.~\cite{brinkmann05}
for wavelengths shorter than the microscopic width of the CL region
is outside the range of the macroscopic capillary model employed,
and thus not physical. Our analysis within the framework of the interface
displacement model generalizes the result of ref.~\cite{dobbs99}
and shows that, at least for small contact angles, negative CL tensions
do not lead to short wavelength instabilities.

\acknowledgments

S. M. thanks S. V. Meshkov and L. Schimmele for fruitful discussions.
M. R. acknowledges financial support from the priority program SPP~1164
{}``Micro and Nano Fluidics'' of the Deutsche Forschungsgemeinschaft
under grant number RA~1061/2-1.

\end{document}